\documentclass[11pt]{article}
\usepackage{hyperref}
\pdfoutput=1
\begin{document}
\title{Volumetric 3-component velocimetry measurements of the flow around a Rushton turbine: A fluid dynamics video}
\author{Kendra Sharp, David Hill, Geoffrey Walters \\ Penn State University \\University Park, PA, 16802, USA \\ \\
Daniel Troolin and Wing Lai \\ TSI Incorporated \\ Shoreview, MN, 55126, USA}
\maketitle
\begin{abstract}
Volumetric 3-component velocimetry measurements have been taken of the flow field near a Rushton turbine in a stirred tank reactor.  This highly unsteady, three-dimensional flowfield is characterized by a strong radial jet, large tank-scale ring vortices, and small-scale blade tip vortices. Approximately 15,000 3d vectors were obtained in a cubic volume; these data offer the most comprehensive view to date of this flow field, especially since they are acquired at three Reynolds numbers (15,000, 107,000, and 137,000). The fluid dynamics video shows various animations and combinations of the velocity and vorticity data. The volumetric nature of the data enable tip vortex identification, vortex trajectory analysis, and calculation of vortex strength. In the video, three identification methods for the vortices are compared based on: the calculation of circumferential vorticity; the calculation of local pressure minima via an eigenvalue approach  (`$\lambda_2$') ; and the calculation of swirling strength again via an eigenvalue approach (`$\lambda_{ci}$'). A `swirl strength' criterion overall provides clearest identification of the tip vortices. The visualization of tip vortices up to 140 degrees past blade passage in the highest Reynolds number case is notable and has not previously been shown.

The associated video can be viewed in \href{http://ecommons.library.cornell.edu/bitstream/1813/13864/4/Sharpetal_APSDFD_mixer09_mp2.m2v}{mpeg-2 format} and \href{http://ecommons.library.cornell.edu/bitstream/1813/13864/3/Sharpetal_APSDFD_mixer09.m1v}{mpeg-1 format}. The related article currently appears in the online first section of Experiments in Fluids (2009), doi 10.1007/s00348-009-0711-9.
\end{abstract}
\end{document}